\documentclass[pra,twocolumn,nobalancelastpage]{revtex4-1}

\usepackage{graphicx,color,amsmath,amsfonts,amssymb,mathtools}

\DeclarePairedDelimiter\bra{\langle}{\rvert}
\DeclarePairedDelimiter\ket{\lvert}{\rangle}
\DeclarePairedDelimiter\rbra{(}{\rvert}
\DeclarePairedDelimiter\rket{\lvert}{)}
\DeclarePairedDelimiterX\braket[2]{\langle}{\rangle}{#1 \delimsize\vert #2}
\DeclareMathOperator{\tr}{tr}
\DeclareMathOperator{\supp}{supp}

\begin{document}

\title{Long range order and symmetry breaking in Projected Entangled Pair State models}

\author{Manuel Rispler}
\author{Kasper Duivenvoorden}
\author{Norbert Schuch}
\affiliation{JARA Institute for Quantum Information,
RWTH Aachen University, D-52056 Aachen, Germany}

\begin{abstract}
Projected Entangled Pair States (PEPS) provide a framework for the
construction of models where a single tensor gives rise to
both Hamiltonian and ground state wavefunction on the same footing. A key
problem is to characterize the behavior which emerges in the system in
terms of the properties of the tensor, and thus of the Hamiltonian.  In
this paper, we consider PEPS models with $\mathbb Z_2$ on-site symmetry and
study the occurence of long-range order and spontaneous symmetry breaking.
We show how long-range order is connected to a degeneracy in the spectrum
of the PEPS transfer operator, and how the latter gives rise to
spontaneous symmetry breaking under perturbations.  We provide a succinct
characterization of the symmetry broken states in terms of the PEPS
tensor, and find
that using the symmetry broken states we can derive a local entanglement
Hamiltonian, thereby restoring locality
of the entanglement Hamiltonian for all gapped phases.  \end{abstract}

\maketitle

\section{Introduction}

\begin{figure}[b]
\includegraphics[width=\columnwidth]{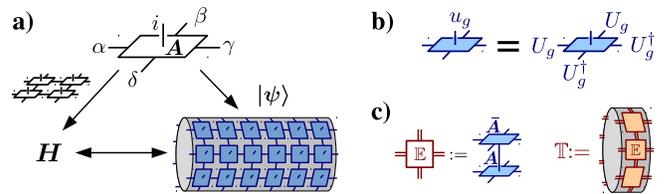}
\caption{
\label{fig:peps}
\textbf{(a)} PEPS models use a tensor $A$ to provide a
description of a wavefunction $\ket\psi$ and an associated parent
Hamiltonian $H$ on the same footing. \textbf{(b)} Properties of the tensor
such as on-site symmetries are inherited by $H$. \textbf{(c)}~The transfer
operator $\mathbb{T}$ encodes all correlations of the wavefunction.}
\end{figure}

Correlated quantum many-body systems exhibit a wide range of
unconventional phenomena.  Their rich physics emerges from the
intricate entanglement structure of those systems, which however at the
same time renders their theoretical study a challenging task. Some of the
most important insights into the physics of those systems
have thus been obtained through wavefunction ansatzes, such as the BCS or
the Laughlin state.  In recent years, ideas from quantum information
have led to classes of ansatz wavefunctions
constructed to capture the entanglement structure
present in interacting quantum
systems~\cite{schollwoeck:review-annphys,verstraete:2D-dmrg,vidal:mera,corboz:fMERA,kraus:fPEPS,pineda:fMERA}.
In particular, Projected Entangled Pair States
(PEPS)~\cite{verstraete:2D-dmrg} describe correlated many-body systems by
associating a local tensor to each site which builds up the 
entanglement in the wavefunction through auxiliary indices.  PEPS
provide a faithful approximation for low-energy states of systems with
local interactions~\cite{hastings:locally,molnar:thermal-peps}, making
them the basis of powerful variational
algorithms~\cite{corboz:fPEPS-sim,orus:tn-review}. At the same time, they
form a framework for the construction of models: From the local
tensor, one can construct parent Hamiltonians which inherit its symmetry
structure and have a global wavefunction built from the tensor as their
ground state~\cite{perez-garcia:parent-ham-2d,schuch:peps-sym}.  Thus,
they can be used to construct ``PEPS models''
(Fig.~\ref{fig:peps}a), where the physics is encoded in a tensor $A$
which gives rise to a local Hamiltonian $H$ and a ground state
wavefunction $\ket\psi$ on the same footing, forming a versatile framework
for the study of correlated quantum
systems~\cite{affleck:aklt-prl,affleck:aklt-cmp,fannes:FCS,rvb-merged,wang:rvb-square-lattice,poilblanc:rvb-superconductors,iqbal:semionic-rvb,chiral-merged,huang:z3-peps-model}.
Within this framework, the key problem is to understand how the local
properties of the tensor (and thus the Hamiltonian) determine the global
properties, i.e., quantum order, of the wavefunction. 

In the last years, considerable progress has been made in the study of
topological order~\cite{wen:book} in PEPS models: It has been understood
how topological order is related to the structure of the local tensor, and
how this allows to construct \emph{all} topological ground states from a
\emph{single} 
tensor~\cite{schuch:peps-sym,buerschaper:twisted-injectivity,sahinoglu:mpo-injectivity}.
More recently, the mechanism behind topological phase transitions within
this framework has been clarified~\cite{schuch:topo-top}, and it has been
found that it can be related to symmetry breaking in the so-called
transfer operator~\cite{haegeman:shadows}.  Finally, PEPS models have also
been shown to provide a natural framework to study the entanglement
properties of correlated quantum systems through entanglement
Hamiltonians~\cite{li:es-qhe-sphere} associated to the boundary
of the system~\cite{cirac:peps-boundaries}.

The aim of this paper is to initiate the development of an analogous
framework for studying conventional long-range order and spontaneous
symmetry breaking within PEPS models, thereby complementing the
understanding of topological phases.  We consider PEPS models
where the local tensor carries a $\mathbb Z_2$ on-site symmetry, and
investigate \emph{i)} under which conditions long-range order emerges, and
\emph{ii)} how the different symmetry broken states can be described in
terms of the single PEPS tensor.  We first show that long-range order
originates in a degeneracy of the transfer operator, and how the latter in
turn leads to spontaneous breaking of the symmetry under fluctuations.  We
subsequently characterize the symmetry broken states in terms of the PEPS
tensor, and show that they correspond to the extreme points of the
degenerate fixed point space of the transfer operator.  We demonstrate
that the symmetry broken PEPS wavefunctions can be used to determine the
spontaneous magnetization, and that they give rise to a local entanglement
Hamiltonian, in contrast to previous
findings~\cite{cirac:peps-boundaries}, which establishes that all gapped
phases can be assigned a short-ranged entanglement Hamiltonian.

\section{
\label{sec:peps-models}
Projected Entangled Pair State models}

Let us start by introducing PEPS.  W.l.o.g., we restrict to a square
lattice with $N_h\times N_v$ sites. A PEPS model is described by a
$5$-index tensor $A^i_{\alpha\beta\gamma\delta}$, $i=0,\dots,d-1$,
$\alpha,\beta,\gamma,\delta=0,\dots,D-1$, with $d$ the physical dimension 
at each site and $D$ the \emph{bond dimension}.  The PEPS
wavefunction $\ket\Psi = \sum c_{i_1\dots i_N}\ket{i_1,\dots,i_N}$ is
obtained by arranging the tensors on the lattice  and contracting
the \emph{virtual indices} $\alpha,\dots,\delta$ of adjacent tensors
as indicated by lines in Fig.~\ref{fig:peps}a, yielding 
$c_{i_1\dots i_N}$.  We will consider systems on a long cylinder (or
torus) of circumference $N_v$, such that the dependence on the boundaries
becomes negligible. 

To any PEPS, one can construct so-called \emph{parent Hamiltonians} which
have the corresponding PEPS as its ground state.  Parent Hamiltonians
ensure that the wavefunction on a small (e.g., $2\times 2$) patch looks
correct, i.e., as being built from the tensor $A$, which is e.g.\
accomplished by a projector onto the span of the tensors on that patch;
under generic conditions,  these
Hamiltonians have a unique finite-volume ground
state~\cite{perez-garcia:parent-ham-2d}.
  By blocking all $N_v$ tensors in a column, a PEPS can equally be
considered as a quasi-1D tensor network (a Matrix Product State, MPS) with
tensors $B^i$ and bond dimension $D^{N_v}$.  In MPS, a central role is
played by the transfer operator $\mathbb{T}=\sum_i B^i\otimes \bar B^i$; in
particular, $\mathbb{T}^{\ell}$ appears in correlation functions at
distance $\ell$, whose decay is therefore governed by the spectrum of
$\mathbb{T}$.  For PEPS, $\mathbb{T}$ has itself a 1D structure,
cf.~Fig.~\ref{fig:peps}c. This leads to a more complex behavior, since in
the thermodynamic limit the system grows simultaneously in both
directions, and thus, the dimension of the space on which $\mathbb T$ acts
grows exponentially with the system size.

Let us now turn towards systems with on-site symmetries.  In PEPS models,
symmetries are encoded locally in the tensor: A symmetry action
$u_g$ on the physical level translates to an action $U_g$ on the
virtual system in a way where it cancels out when
contracting tensors, see Fig.~\ref{fig:peps}c, giving rise to an
invariant wavefunction $\ket\Psi = U_g^{\otimes N}
\ket\Psi$~\cite{perez-garcia:inj-peps-syms}.  
This induces a symmetry 
$[\mathbb{T},U_g^{\otimes N_v}\otimes\bar{U}_g^{\otimes N_v}]=0$
of $\mathbb{T}$, which is therefore block-diagonal in a basis of
irreducible representations (irreps) of $U_g$. At the same time, the parent
Hamiltonian $H$ enjoys by construction the same symmetry, i.e., we obtain
a PEPS model with symmetry $[H,u_g^{\otimes N}]=0$.  An instructive
example is the ``Ising
PEPS''~\cite{verstraete:comp-power-of-peps} with
$A=\ket0\bra{\theta,\theta,\theta,\theta}+
\ket1\bra{ \bar\theta, \bar\theta, \bar\theta, \bar\theta}$, where the ket
(bra) corresponds to the physical (virtual) indices, and $\ket\theta =
\cos\theta\,\ket0+\sin\theta\,\ket1$, 
$\ket{\bar\theta}= \sin\theta\,\ket0+\cos\theta\,\ket1$.
This model has a $\mathbb{Z}_2$
symmetry with non-trivial action $u_1=U_1=X=\left(\begin{smallmatrix}
0&1\\1&0 \end{smallmatrix}\right)$. The wavefunction is of the form
\[
\ket\Psi = \sum_{i_1,\dots, i_N}
 e^{-\beta/2\, H_\mathrm{cl}(i_1,\dots,i_N)}\ket{i_1,\dots,i_N}
\]
with
$H_\mathrm{cl}$ the classical 2D Ising model; it thus
has the same $\sigma_z$ correlation functions 
and therefore undergoes a second-order phase transition 
at 
\begin{equation}
\label{eq:ising-crit}
\theta_c=\tfrac12\arcsin\Big[\tfrac{1}{\sqrt{1+\sqrt{2}}}\Big]\approx 0.349596\ .
\end{equation}

\section{Long-range order and spontaneous symmetry breaking}

As we have seen, PEPS form a natural framework to model systems with
on-site symmetries: By encoding the symmetry locally into the tensor, we
obtain a local Hamiltonian with the same symmetry which depends
smoothly on the parameters of the tensor, and whose ground state
wavefunction can be constructed from the very same tensor. However, we
have also found that by changing parameters, such a model can undergo a phase
transition to a symmetry broken phase.  Nevertheless, the PEPS
wavefunction $\ket\Psi$, which is the unique ground state of the system,
remains invariant under the symmetry throughout the phase diagram,
$\ket\Psi=u_g^{\otimes N}\ket\Psi$; e.g., for the Ising PEPS at zero
temperature ($\theta=0$), we have
$\ket\Psi=\tfrac{1}{\sqrt{2}}(\ket{0\cdots0}+\ket{1\cdots1})$.  This
leads to the central question of this work: How can we characterize symmetry
breaking in PEPS models, and how can we construct the distinct symmetry
broken states starting from a single \emph{symmetric} tensor $A$?

Let us for a moment leave aside PEPS models and consider the analogous
question for general Hamiltonians with a symmetry: Generically, these
systems have a unique finite volume ground state which therefore
carries the full symmetry of the Hamiltonian.  Since the average
magnetization $\big\langle\tfrac1N \sum_i Z_i\big\rangle$ in those states
is zero, the ordered phase is characterized through non-vanishing
long-range order $\big\langle \tfrac{1}{N^2}\sum_{ij} Z_i
Z_j\big\rangle\sim \mathrm{const}\,$.  In order to obtain the symmetry
broken states, one couples the system to a small external field
$H'=H+h\sum_i Z_i$ and considers the ground state $\ket{\Psi_{N,h}}$,
taking first the limit $N\to\infty$ and subsequently $h\to0$; if in the
limit $\big\langle \tfrac1N \sum_i Z_i\big\rangle\ne0$, one says that the
system exhibits spontaneous symmetry breaking. 

How are these two notions of long-range order and
spontaneous symmetry breaking related, and can one construct an
approximate symmetry broken state for a finite system?
Given a unique finite-volume ground state with
long-range order such as $\ket\Psi\approx\ket{0\cdots0}+\ket{1\cdots1}$ and
order parameter $Z$, one can construct an orthogonal
state $\ket\Phi\propto\sum_i Z_i\ket\Psi$ whose energy approaches the
ground state as $N\rightarrow\infty$~\cite{horsch:lowlying-state}. A
generic local perturbation will select a unique ground state from the
space spanned by $\ket\Psi$ and $\ket\Phi$, assuming a gap above.  In
fact, it can be proven that the only states which do not exhibit
long-range order  for any observable $O$ with $\langle O\rangle=0$,
i.e., which are stable under arbitrary perturbations and thus form the
symmetry broken states, are
$\ket\Psi\pm\ket\Phi$~\cite{koma:symbreaking-finitesize}; this is, 
the symmetry broken states can be constucted from the symmetric ground
state $\ket\Psi$ alone.  It is straightforward to see from this argument
that the value of the long-range order is just the spontaneous
magnetization squared.  It should be pointed out, however, that the above
argument relies on the validity of the perturbative treatment which cannot
be justified rigorously in the limit $N\to\infty$, as the
\emph{total} perturbation diverges, and in fact, the equality between the
two notions of symmetry breaking and the values of the corresponding order
parameters has not been proven rigorously except for very few cases such
as the Ising model~\cite{simon:book-lattice-gases}.

In the following, we will follow a very similar reasoning in order to
relate long-range order and spontaneous symmetry breaking in PEPS, and in
particular to show how to construct the set of symmetry-broken states
(i.e., those which are stable under arbitrary local perturbations)
starting from a single object, namely the unique symmetric PEPS
wavefunction $\ket\Psi$ and its associated symmetric tensor $A$. Here, a
central role will be played by the one-dimensional transfer operator and
in particular its spectral properties and eigenstates, partly taking the
role of the Hamiltonian in the preceding discussion.  We will start by
showing that long-range order in the wavefunction is closely related to an
approximate degeneracy in the spectrum of the transfer operator.
Subsequently, we study the behavior of the transfer operator under
perturbations, and we show that \emph{arbitrary} perturbations induce a
splitting of the degenerate subspace in a \emph{fixed} basis---independent
of the perturbation---which therefore describes the symmetry broken states
of the model.

In the following discussion, we will restrict to the symmetry group
$\mathbb{Z}_2$, and denote the physical (virtual) symmetry action by $x$
($X$).  

\subsection{Long-range order and the spectrum of the transfer operator}

Let us first see how long-range order in a PEPS model with $\mathbb Z_2$
on-site symmetry is related to the properties of the underlying tensor and
in particular the transfer operator. We start by defining long-range order
(in analogy to Ref.~\cite{koma:symbreaking-finitesize}):
We say that $\ket\Psi$ has long-range order if there exists a 
 local operator $Z=Z^\dagger$ with $Zx=-xZ$ and $\|Z\|_\mathrm{op}\le1$
(the \emph{order parameter}), and $c>0$ s.th.\ for sufficiently
large $N_v$, 
\begin{equation}
\label{eq:lro}
\lim_{N_h\to\infty}\frac{1}{N_hN_v^2}\sum_{m,n} \bra\Psi Z_m Z_n\ket\Psi
\ge c N_v 
\end{equation}
where the sum runs over all sites---i.e., on an infinite cylinder, the
spins are correlated at least over a distance proportional to the
circumference. [Away from fixed point wavefunctions, we indeed cannot
expect correlations along the cylinder over arbitrary distances; on the
other hand, e.g.\ in the Ising PEPS correlations only break down after a
distance $\exp(cN_v)$.] Note that $Z$ can depend on the lattice site, such
as in the case of an antiferromagnet.  In such a case, we will also have
to restrict the possible values of $N_v$ accordingly, e.g.\ to even
numbers (or we have to block tensors),
and we will tacitly assume this in the following.  (Corresponding
restrictions on $N_h$ will not be necessary in the limit $N_h\to\infty$
away from fixed point wavefunctions.)

In 1D, i.e., MPS, long-range order is in one-to-one correspondence with a
degeneracy of the transfer operator $\mathbb{T}$---roughly speaking, the
only way to build long-range correlations over arbitrary distances using
the constant virtual dimension is an exact degeneracy of the leading
eigenvalue of $\mathbb T$.
In 2D, however, the situation is much less clear due
to the exponentially growing dimension of the virtual space on which
$\mathbb{T}$ acts. However, as we will show, under certain conditions
Eq.~(\ref{eq:lro}) 
still implies the existence of an almost degenerate second eigenvalue. 
To this end, define a ``dressed'' transfer operator $\mathbb{T}_{Z_s}$,
where we insert $Z$ on the physical index at position $s=1,\dots,N_v$, and
correspondingly $\mathbb{T}_{\hat
Z}=\tfrac{1}{N_v}\sum_{s=1}^{N_v}\mathbb{T}_{Z_s}$; such operators arise
in expectation values such as Eq.~(\ref{eq:lro}).  Assume that
$\mathbb{T}$ is diagonalizable, $\mathbb{T}=\sum \lambda_i \lvert
r_i)(l_i\rvert$ with $(l_i\vert r_j)=\delta_{ij}$.  Here, we use $|\cdot)$ to
denote (eigen-)vectors on the level of the transfer operator, i.e., on the
virtual indices;  due to the
ket-bra structure of the transfer operator, these vectors can themselves
be regarded as operators, in which case we will write them without
brackets. Since $[\mathbb T,X^{\otimes N_v}\otimes X^{\otimes N_v}]=0$,
the eigenvectors of $\mathbb T$ transform even or odd under the symmetry
(i.e., as different irreps),
and we will denote the largest eigenvalue in the even (odd) symmetry
sector by $\lambda_+$ ($\lambda_-$), with corresponding eigenvectors
$\rket{r_\pm}\rbra{l_\pm}$.
We thus have that $(X^{\otimes N_v}\otimes
X^{\otimes N_v})\rket{r_\pm}=\pm\rket{r_\pm}$, or (if interpreting the
eigenvectors as operators)
\begin{equation}
\label{eq:eigvecsym}
X^{\otimes N_v}r_\pm X^{\otimes N_v} = \pm r_{\pm}\ .
\end{equation}
Note that this implies that $r_-$ cannot be positive semidefinite, and
since $\mathbb T$ must have a positive semidefinite fixed point (as it is
a completely positive map),  it follows that $|\lambda_+|\ge|\lambda_-|$.

Our goal is to show that $\lambda_+$ and $\lambda_-$, as a function of
$N_v$, become degenerate as $N_v\rightarrow\infty$. We can thus assume
that $|\lambda_-|<|\lambda_+|$ (otherwise there is nothing to show).  We
will additionally restrict to the case where the largest eigenvalue
$\lambda_+$ is non-degenerate also within the even parity sector, since an
exact degeneracy at \emph{finite} $N_v$ hints an additional symmetry in
the wavefunction which we would have to incorporate for a full
description.  This implies
$\lambda_+>0$~\cite{perez-garcia:mps-reps,evans:cpmap-positive-eigenvalue}
and we can w.l.o.g.\ normalize our tensors such that $\lambda_+=1$.  We
can now re-express Eq.~(\ref{eq:lro}) as 
\begin{equation}
\label{eq:lro-top}
c N_v \le \lim_{N_h\rightarrow\infty}
    \sum_{p=0}^{N_h-2}\frac{\tr[\mathbb{T}_{\hat Z}\mathbb{T}^p
		\mathbb T_{\hat Z} \mathbb T^{N_h-p-2}]}{
	\tr[\mathbb T^{N_h}]} + 1 \ ,
\end{equation}
where the additive term $+1$ stems from the case where both $Z$'s
are in the same column, and using $\|Z\|_\mathrm{op}\le 1$. In order
to simplify the r.h.s., we split the sum at $p=N_h/2-1$ and use cyclicity of
the trace, together with $p\leftrightarrow N_h-p-2$, to obtain two
identical sums,
\begin{equation*}
c N_v \le \lim_{N_h\rightarrow\infty}
    2\sum_{p=0}^{N_h/2-1}\frac{\tr[\mathbb{T}_{\hat Z}\mathbb{T}^p
		\mathbb T_{\hat Z} \mathbb T^{N_h-p-2}]}{
	\tr[\mathbb T^{N_h}]} + 1\ .
\end{equation*}
(If $N_h$ is even, the term $p=N_h/2-1$ appears only once, but this bears no
relevance.) Since the largest eigenvalue of $\mathbb T$ is non-degenerate,
we have that
\begin{equation}
\label{eq:top-exp-conv}
\|\mathbb T^M-|r_+)(l_+|\|_\mathrm{tr}\le c \Gamma^M\ ,
\end{equation}
where $\Gamma<1$ upper bounds the second largest eigenvalue
of $\mathbb T$~\cite{szehr:channel-convergence-blaschke}. (Note though
that $c$ can heavily depend on $N_v$ and properties of $\mathbb T$.) Using
a sequence of triangle inequalities and taking the limit $N_h\to\infty$
(see Appendix for details), we can then replace $\mathbb
T^{N_h}\to|r_+)(l_+|$ and obtain
\begin{align}
c N_v &\le 2\sum_{p=0}^\infty \tr\big[
	\mathbb T_{\hat Z} \mathbb T^p \mathbb T_{\hat Z}
	\rket{r_+}\rbra{l_+}\big] + 1
\label{eq:lro-top-fpt}
\\
&=2\sum_i \sum_{p=0}^\infty \lambda_i^p
    \underbrace{\rbra{l_+} \mathbb T_{\hat Z}\rket{r_i}
    \rbra{l_i} \mathbb T_{\hat Z}\rket{r_+}}_{=:m_i}+1\ ,
\nonumber
\end{align}
where in the last step, we have expanded $\mathbb T$ in its eigenbasis.

As $\mathbb{T}_{\hat Z}$ anticommutes with the symmetry, 
$(l_+|\mathbb T_{\hat Z}|r_i)=0$ for eigenvectors $|r_i)$ from the even
sector. Thus, only eigenvectors from the odd sector contribute to the sum.
For those eigenvectors, $|\lambda_i|\le|\lambda_-|$, and therefore
\[
\left|\sum_p \lambda_i^p m_i\right|\le 
\frac{1}{1-|\lambda_i|}|m_i|\le\frac{1}{1-|\lambda_-|}|m_i|\ .
\]
If now $m_i\ge0$,
we have $\sum_i
|m_i|=\rbra{l_+}\mathbb{T}_{\hat{Z}}\mathbb{T}_{\hat{Z}}\rket{r_+}\le 1$
(since $\|Z\|_\mathrm{op}\le 1$)
and thus \[
c N_v \le \frac{2}{1-|\lambda_-|}+1\ ,
\]
i.e., $|\lambda_-| \ge 1-O(1/N_v)$: We find that long-range order
implies that the gap of $\mathbb{T}$ closes in the thermodynamic limit at
least as $1/N_v$.
The required condition $m_i\ge0$ is automatically satisfied if
$\mathbb{T}$ and $\mathbb{T}_{\hat Z}$ are hermitian, which in particular
holds if the tensor and thus the PEPS model is invariant under combined
reflection and time reversal, but turns out to be true also for a range of
other models~\footnote{ Note that even an exponential lower bound $b^{N_v}$ 
on the r.h.s.\ of Eq.~(\ref{eq:lro}) might not be strong enough to bound the
gap without extra assumptions: The convergence of $\mathbb{T}^L$ to its
fixed point can be roughly bounded by $\gamma^{L} L^{D^{2 N_v}}$, where
$\gamma = |\lambda_{-}/\lambda_+|$ and $D^{2 N_v}$ is the dimension of
the space $\mathbb{T}$ acts on~\cite{szehr:channel-convergence-blaschke}.
Correlations are thus lost at a scale set by $\gamma^{L} \sim
L^{D^{2 N_v}}$;  in order for a lower bound $b^{N_v}$ to 
imply a bound on $\gamma$ we thus need $b>D^2$.}.

We thus find that under certain conditions, long-range order implies that
the transfer operator has approximately degenerate eigenvalues
$\lambda_+\approx \lambda_-$ in the even and odd sector, together
with non-vanishing matrix elements $\rbra{l_\pm}\mathbb{T}_{\hat
Z}\rket{r_\mp}$.
Conversely, an approximately degenerate eigenvalue in the odd sector,
together with non-vanishing matrix elements, will clearly give rise to
long-range order.

\subsection{Spontaneous symmetry breaking}

Let us now investigate spontaneous symmetry breaking in PEPS models with
long-range order, and in particular how to construct the symmetry broken
states.  The symmetry broken states are those states which remain ground
states under generic perturbations, where for a given perturbation, we
first need to take the thermodynamic limit and subsequently take the
strength of the perturbation to zero.  In the context of PEPS models, the
natural perturbations to consider are perturbations of the tensor which
can be realized by acting solely on the physical index,
$A^i_{\alpha\beta\gamma\delta} \to \sum(\openone+\Lambda)_{ij}
A^j_{\alpha\beta\gamma\delta}$ with $\|\Lambda\|\ll 1$, as these both
correspond to small perturbations of the parent Hamiltonian and leave us
in the PEPS manifold~\cite{cirac:itb}. Such a perturbation gives in turn
rise to a perturbation of the transfer operator,
$\mathbb{T}\to\mathbb{T}_{[\Lambda]}$, and thus the symmetry broken states
will exactly correspond to boundary conditions given by the stable
fixed points of $\mathbb{T}_{[\Lambda]}$ in the appropriate limit.

\subsubsection{Hermitian transfer operator}

How will the transfer operator of a system with long-range order respond
to such perturbations?  To this end, consider the limit
$N_v\rightarrow\infty$ where $\lambda_-/\lambda_+\rightarrow 1$, and let
us assume that $\mathbb{T}$ has a gap below the two degenerate
eigenvalues, $\mathbb{T}=
\rket{r_+}\rbra{l_+}+\rket{r_-}\rbra{l_-}+\dots$.  A sufficiently small
perturbation will then induce a splitting within this subspace without
mixing it with lower-lying eigenvectors; the basis in which this splitting
occurs is independent of the strength of $\Lambda$ 
 and thus leaves us
with a single fixed point $\mathbb{T}_{[\Lambda]}^{N_h}\sim
\rket{r_\uparrow}\rbra{l_\uparrow}$ in the limit $N_h\to\infty$, even if
later $\|\Lambda\|\to0$.  This fixed point is therefore the boundary
condition for the symmetry broken state under perturbation $\Lambda$. 

Naively, one might expect that the symmetry broken fixed point sensitively
depends on the exact form of the perturbation chosen.  However, as we will
show in the following, the symmetry broken fixed point
$|r_\uparrow)(l_\uparrow|$ is independent of $\Lambda$.  We will first consider 
the case of a hermitian transfer operator, i.e., $\rbra{r_i}=\rbra{l_i}$ and
$(r_i\rket{r_j}=\delta_{ij}$, and a hermiticity-preserving perturbation.
A generic such perturbation to $\mathbb T$ will induce a splitting of the two
degenerate eigenvalues in
an orthogonal basis \begin{align*}
\rket{r_\uparrow}&\propto \rket{r_+} + \gamma\rket{r_-}\\
\rket{r_\downarrow}&\propto \rket{r_+} - \frac{1}{\gamma^*}\rket{r_-}\ .
\end{align*}
Since $r_{\uparrow\downarrow}$ are non-degenerate eigenvectors of $\mathbb
T_{[\Lambda]}$, which is a completely positive (CP) map, they must be
hermitian, and thus $\gamma\in\mathbb R$.
Since $\mathbb{T}_{[\Lambda]}$ is a completely positive map, its leading
(non-degenerate) eigenvector must be a positive operator, 
\begin{equation}
r_\uparrow\propto r_++\gamma r_-\ge0\ .
\end{equation}
Using Eq.~(\ref{eq:eigvecsym}) and the fact that conjugation preserves
positivity, we find that also
\begin{equation}
r_+ -\gamma r_- 
=
X^{\otimes N_v}(r_+ +\gamma r_-)X^{\otimes N_v} \ge 0\ .
\end{equation}
Since for a pair of positive operators $P,Q\ge0$, it holds that
$\tr[PQ]\ge0$, it follows that 
\begin{align*}
0&\le \tr[ (r_++\gamma r_-) (r_+-\gamma r_-)]\\
&= (r_+\rket{r_+} - \gamma^2 (r_-\rket{r_-}\ ,
\end{align*}
and thus
$|\gamma|\le1$.  
By changing the sign of $\Lambda$, we can exchange the two leading
eigenvectors, i.e., $|r_\downarrow)$ becomes the leading eigenvector, and
thus $r_\downarrow\ge0$.  Following the same line of reasoning as before,
this yields $|\gamma|\ge1$, and
thus, $\gamma=\pm1$.  We thus find that regardless of the perturbation,
the transfer operator always acquires the same pair of fixed points
\[
\rket{r_{\uparrow\downarrow}}\propto \rket{r_+}\pm\rket{r_-}\ ,
\]
solely as a consequence of its complete positivity. Note that
$r_{\uparrow\downarrow}$ are the extremal positive states $r_++\gamma
r_-\ge0$, i.e., the stable symmetry broken states are those where the
symmetry is maximally broken, as intuitively expected.

\subsubsection{Non-hermitian transfer operator}

Let us now consider the non-hermitian case.  Since
$|r_+)(l_+|+|r_-)(l_-|=|r_\uparrow)(l_\uparrow|+|r_\downarrow)(l_\downarrow|$,
it follows that 
\[
|r_{\uparrow\downarrow})\propto|r_+)\pm\lambda|r_-) \mbox{\ and\ }
(l_{\uparrow\downarrow}|\propto(l_+|\pm\tfrac1\lambda(l_-|
\]
where hermiticity of the eigenvectors implies $\lambda\in\mathbb R$, and
we w.l.o.g. choose $\lambda>0$. Again, complete positivity of 
$\mathbb T_{[\Lambda]}$ and $\mathbb T^\dagger_{[\Lambda]}$,
 together with the possibility to change the
ordering of eigenvectors by changing the sign of $\Lambda$, implies that
$r_{\uparrow\downarrow},l_{\uparrow\downarrow}\ge0$.
To determine
$\lambda$, consider 
\begin{equation}
\label{eq:S-def}
\mathcal S:=\mathrm{supp}\,r_+\equiv(\mathrm{ker}\,r_+)^\perp\ ,
\end{equation}
and let $\Pi_\mathcal S$ be the orthogonal projector onto $\mathcal S$.
Complete positivity of $\mathbb T$ implies that
$\mathrm{supp}\,r_-\subset\mathcal S$~\footnote{Otherwise, we could find a
$P\ge0$ s.th.\ $(P|r_+)=0$, but $(P|r_-)\ne0$. Applying $\mathbb T$
repeatedly to $P\ge0$ would therefore converge to the non-positive $l_-$,
$(P|\mathbb{T}^\infty=(l_-|$, which would be in contradiction to $\mathbb
T$ being completely positive, i.e., mapping positive operators to positive
operators.}, and therefore
$\mathrm{supp}\,r_{\uparrow\downarrow}\subset\mathcal{S}$, i.e.,
$\Pi_\mathcal{S}r_{\uparrow\downarrow}\Pi_{\mathcal{S}}=r_{\uparrow\downarrow}$.
Define
\[
\tilde{l}_{\uparrow\downarrow}=\Pi_{\mathcal{S}}l_{\uparrow\downarrow}
\Pi_{\mathcal{S}}\ge0\ ;
\] 
since 
$(\tilde{l}_{\uparrow\downarrow}|r_{\uparrow\downarrow})=
\tr[\tilde{l}^\dagger_{\uparrow\downarrow}r_{\uparrow\downarrow}]
=
\tr[\Pi_{\mathcal S} l^\dagger_{\uparrow\downarrow}\Pi_{\mathcal
S}r_{\uparrow\downarrow}]=
\tr[l^\dagger_{\uparrow\downarrow} r_{\uparrow\downarrow}]=
(l_{\uparrow\downarrow}|r_{\uparrow\downarrow})=1$,
we have that 
\begin{equation}
\label{eq:ltilde-ne-0}
\tilde l_{\uparrow\downarrow}\ne 0\ .
\end{equation}
  As $r_{\uparrow\downarrow}=r_+\pm\lambda
r_-$, and $r_+\ge0$, $\lambda\ge0$, it follows that 
$\ker r_\uparrow\cap \mathcal S$ ($\ker
r_\downarrow\cap \mathcal S$) is contained in the negative (positive)
eigenspace of $r_-$, which implies that 
\begin{equation}
\label{eq:ker-rup-rdn-orth}
(\ker r_\uparrow\cap \mathcal
S)\perp(\ker r_\downarrow\cap\mathcal S)\ .
\end{equation}
Moreover, 
since $\tr[\tilde{l}_{\uparrow\downarrow}r_{\downarrow\uparrow}]=
\tr[l_{\uparrow\downarrow}r_{\downarrow\uparrow}]=
(l_{\uparrow\downarrow}|r_{\downarrow\uparrow})=0$, we have $\supp
\tilde{l}_{\uparrow\downarrow}\subset\ker r_{\downarrow\uparrow}$, and
thus 
\begin{equation}
\label{eq:supp-ker-S}
\supp\tilde{l}_{\uparrow\downarrow}\subset\ker
r_{\downarrow\uparrow}\cap\mathcal{S}
\end{equation}
(as $\supp\tilde{l}_{\uparrow\downarrow}\subset\mathcal{S}$).
Eq.~(\ref{eq:supp-ker-S}) has two implications:
First, together with Eq.~(\ref{eq:ltilde-ne-0}),
$\tilde{l}_{\uparrow\downarrow}\ne0$,
it shows that $\ker r_{\uparrow\downarrow}\cap\mathcal{S}\ne0$. Since 
$\mathcal{S}=\supp r_+$ [Eq.~(\ref{eq:S-def})],
this implies that $r_{\uparrow\downarrow}$ are again the extremal points
of the positive cone $r_++\lambda r_-\ge0$. 
Second, by combining Eq.~(\ref{eq:supp-ker-S}) with
Eq.~(\ref{eq:ker-rup-rdn-orth}), we find that 
$\supp \tilde{l}_{\uparrow} \perp\supp
\tilde{l}_{\downarrow}$, and thus 
\[
0=\tr[\tilde l_{\uparrow}\tilde
l_{\downarrow}]
=\tr[\Pi_\mathcal S(l_+ + \lambda l_-)\Pi_\mathcal S(l_+-\lambda l_-)]\ ,
\] which allows us to determine the corresponding value of
$\lambda$ as $\lambda = \big[\tr[(\Pi_{\mathcal S}l_+)^2]/
\tr[(\Pi_{\mathcal S}l_-)^2])\big]^{1/2}$. Together, this proves that also in the non-hermitian
case, the fixed points of the transfer operator under perturbations, i.e.,
the symmetry broken states, are uniquely determined independent of
$\Lambda$, and correspond to the extremal symmetry-broken states.

\section{Numerical study and entanglement Hamiltonians}

Having understood the structure of the symmetry broken states in the
thermodynamic limit, let us now investigate the accuracy of our result
for finite systems, and use our findings to re-examine the entanglement
spectra and Hamiltonians of models with symmetry breaking.  

To this end, we have performed numerical simulations for two models.
First, we have considered the Ising PEPS introduced in
Sec.~\ref{sec:peps-models}.  As its transfer
operator is isometric to the transfer operator of the 2D classical Ising
model and it exhibits the same $\sigma_z$ correlation functions, it
allows us to benchmark our numerical findings against exact results.

As a second model, we have studied the square lattice AKLT model with
nematic field~\cite{cirac:peps-boundaries}. The AKLT
model~\cite{affleck:aklt-cmp} is constructed by placing spin-$\tfrac12$
singlets on the links and subsequently projecting them onto the spin $2$
subspace, i.e., $A$ is of the form
$\Pi_{S=2}(\openone\otimes\openone\otimes \sigma_y\otimes \sigma_y)$,
where $\Pi_{S=2}$ projects onto the spin-$2$ space; the model can be
mapped to a ferromagnetic model through a sublattice rotation.
Subsequently, a ``nematic field'' $A\to \sum
N_{ij}A^j_{\alpha\beta\gamma\delta}$, $N=\exp(\alpha S_z^2)$ favoring
large values of $S_z^2$ is applied to the AKLT tensor which yields a model
with $U(1)\rtimes \mathbb{Z}_2$ symmetry  and eventually leads to breaking
of the $\mathbb{Z}_2$ symmetry~\cite{cirac:peps-boundaries}.

\begin{figure}[t]
\includegraphics[width=0.95\columnwidth]{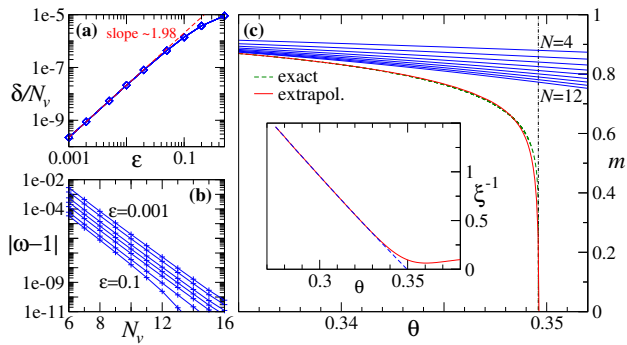}
\caption{
\label{fig:ising-conv-mag}
Numerical results for the Ising PEPS. \textbf{(a,b)} Error analysis for
$\Lambda=\epsilon Z$, $\theta=\pi/16$: a) Scaling of the error
$\delta=1-|(f|g)|^2$ between the fixed point of $\mathbb{T}_{[\Lambda]}$,
$|f)$,  and of its projection onto $|r_\pm)$, $|g)$; we find
$\delta/N_v\sim \epsilon^2$. (The data for different $N_v$
almost exactly
coincide.) b) The deviation of $\omega=|(r_-|g)/(r_+|g)|^2$ from 
$1$ vanishes exponentially in $N_v$.
\textbf{(c)}~Magnetization, computed
from the 
symmetry broken states $|r_+)+|r_-)$, for $N_v=4,\dots,12$ (blue), 
extrapolated curve (red), and  exact solution from the correspondence to
the classical 2D Ising model (dashed green). 
} \end{figure}

In all cases, the simulations have been carried out through an exact
diagonalization of the transfer operator (cf., e.g.,
Ref.~\cite{schuch:rvb-kagome}), where the transfer operator $\mathbb T$ is
applied to a vector by sequential contraction of indices rather than by
building its matrix representation. Thus, the largest objects which need
to be stored are on the order of the size of an eigenvector, $D^{2N_v}$,
rather than that of the transfer operator, $D^{4N_v}$; at the same time,
one obtains results with machine precision, allowing for an accurate
scaling.

\subsection{Error scaling}

\begin{figure}[t]
\includegraphics[width=0.95\columnwidth]{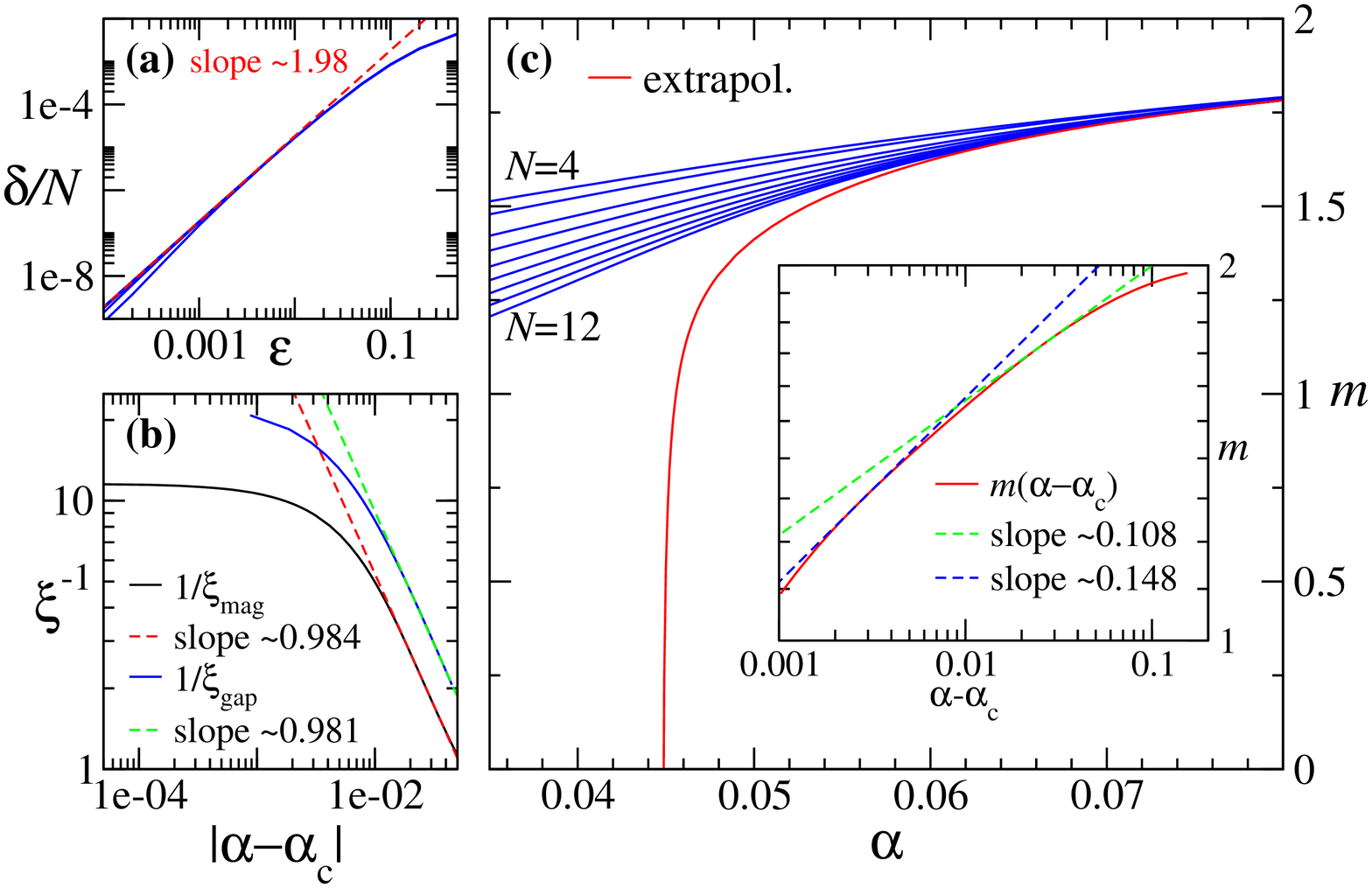}
\caption{
\label{fig:aklt-top-mag}
Nematic AKLT model. \textbf{(a)} Error analysis,
cf.~Fig.~\ref{fig:ising-conv-mag}b.
\textbf{(b)} Critical exponents for the correlation length, 
determined as $\xi_\mathrm{gap}=-1/\ln(\lambda_-/\lambda_+)$ in the
disordered phase and $\xi_\mathrm{mag}\equiv\xi(\theta)$ from the
magnetization fit (cf.~Fig.~\ref{fig:ising-conv-mag}c) in the
symmetry-broken phase. \textbf{(c)} Magnetization $m$ for $N_v=4,\dots,12$ and
extrapolated value (cf.~Fig.~\ref{fig:ising-conv-mag}c); the inset shows
the critical exponent of $m$.
}
\end{figure}

Our derivation involved a number of assumptions, in particular neglecting
the coupling to eigenstates other than $|r_\pm)$ in second and higher
order perturbation theory even in the limit $N_v\to\infty$, and neglecting
the splitting  between $\lambda_{\pm}$~\footnote{%
  Given the notorious difficulty to establish mathematically rigorous
  results about symmetry breaking for most classical
  models~\cite{simon:book-lattice-gases}, and the fact that any classical
  model corresponds to a PEPS model~\cite{verstraete:comp-power-of-peps}, a
  fully rigorous derivation is indeed elusive.}. 
In order to assess the validity of neglecting the coupling to the other
levels, we have compared the fixed point $|f)$ of the perturbed transfer
operator $\mathbb{T}_{[\Lambda]}$, with $\Lambda$ a general perturbation,
to the fixed point $|g)$ obtained after projecting
$\mathbb{T}_{[\Lambda]}$ onto $\rket{r_\pm}$ and $\rbra{l_\pm}$.
Fig.~\ref{fig:ising-conv-mag}a shows the result for the Ising PEPS, where
we find that the error $\delta = 1-|(f|g)|^2$ scales as $\delta\sim N_v
\|\Lambda\|^2$.  This scaling is consistent with a second order
perturbation treatment, where one assumes that the local perturbation
$\Lambda$ has only a local effect on $|r_{\pm})$, and thus can only be
undone by  a term in its vicinity, leading to $O(N_v)$ contributions,
rather than $O(N_v^2)$, of amplitude $\|\Lambda\|^2$ each. The same
behavior is found for the nematical AKLT model,
cf.~Fig.~\ref{fig:aklt-top-mag}a. 
We have subsequently also verified the validity of neglecting the
splitting between $\lambda_{\pm}$: Fig.~\ref{fig:ising-conv-mag}b shows
that the exact solution for the two-dimensional degenerate subspace,
$|g)$, in turn converges exponentially to $|r_{\uparrow\downarrow})$ as
$N_v\to\infty$, consistent with an exponentially vanishing splitting
$1-\lambda_-/\lambda_+$.

\subsection{Magnetization}

The knowledge of the boundary conditions $|r_{\uparrow\downarrow})$ and
$(l_{\uparrow\downarrow}|$ of the symmetry broken states allows us to
compute the approximate spontaneous magnetization for finite $N_v$.   For
the Ising PEPS, we find that the finite $N_v$ data reproduces the
analytic expression for the infinite system very well away from the
critical point, with the error vanishing exponentially in both $N_v$ and
the distance from the critical point; this is consistent with the
expectation that the error should vanish as $e^{-N_v/\xi}$ with $\xi$ the
correlation length. On the other hand, there are strong finite size
effects as we approach the critial point, as shown in
Fig.~\ref{fig:ising-conv-mag}c; indeed, it is natural to expect that finite
size effects will dominate as soon as $\xi\sim N_v$.  Nevertheless, it is still
possible to accurately extrapolate the magnetization curve to the
thermodynamic limit even very close to the critical point. To this end, we
first determine the behavior of the correlation length in the proximity of
the critical point by fitting $m_\theta(N_v)=a+b\exp(-N_v/\xi_\theta)$
as a function of $N_v$, and subsequently fitting $1/\xi_\theta$
quadratically (shown in the inset; note that this makes an implicit
assumption about the critical
exponent of $\xi$); the critical point from the fit is
$\theta_{c,\xi}\approx 0.349666$, as compared to the exact value
$\theta_{c,\mathrm{ex}}\approx 0.349596$.
The resulting fitting function $\xi'(\theta)$ is then
used in the scaling ansatz $m_\theta(N_v)=m_\theta(\infty)+f N_v^{-g}
\exp[-N_v/\xi'(\theta)]$, where the algebraic factor $N_v^{-g}$ accounts
for short-range effects~\footnote{The reason that we fit $\xi'_\theta$
independently in a first step is that in the relevant regime
where $N_v\ll\xi'_\theta$, the second fit does not allow for an accurate
determination of $\xi'(\theta)$ and might even return negative values.}.
Altogether, this extrapolation yields a very accurate approximation to the
analytical curve even in the regime very close to the critical point where
finite size effects seemingly dominate, as shown in
Fig.~\ref{fig:ising-conv-mag}c.

We have subsequently applied the same analysis to determine the
magnetization curve of the nematic AKLT model. The data for
$N_v=4,\dots,12$ and the extrapolated magnetization curve are shown in
Fig.~\ref{fig:aklt-top-mag}c, and we find a value of $\alpha_c\approx 0.0447$
for the critical point. Using this value of $\alpha_c$, we have
subsequently studied the critical scaling of the model.
Fig.~\ref{fig:aklt-top-mag}b shows the critical scaling of two different
correlation lengths---$\xi_\mathrm{mag}$, obtained by fitting the
magnetization as a function of $N_v$ as
$m(N_v)=a+b\exp(-N_v/\xi_\mathrm{mag})$, and $\xi_\mathrm{gap}$, obtained
by an exponential fit $\xi(N_v)=\xi_\mathrm{gap}+a\exp(-bN_v)$ of the
correlation length $\xi(N_v)=-\ln(\lambda_-/\lambda_+)$ extracted from the
gap of the
transfer operator---both of which yield a critical exponent of $\nu\approx1$
for the correlation length. The inset of 
inset of Fig.~\ref{fig:aklt-top-mag}c shows the corresponding analysis for
the magnetization, which is compatible with a critical exponent
$\beta=1/8$, therefore suggesting that the nematic AKLT model is 
in the 2D Ising universality class.

\subsection{Entanglement Hamiltonian}

\begin{figure}[t]
\includegraphics[width=0.85\columnwidth]{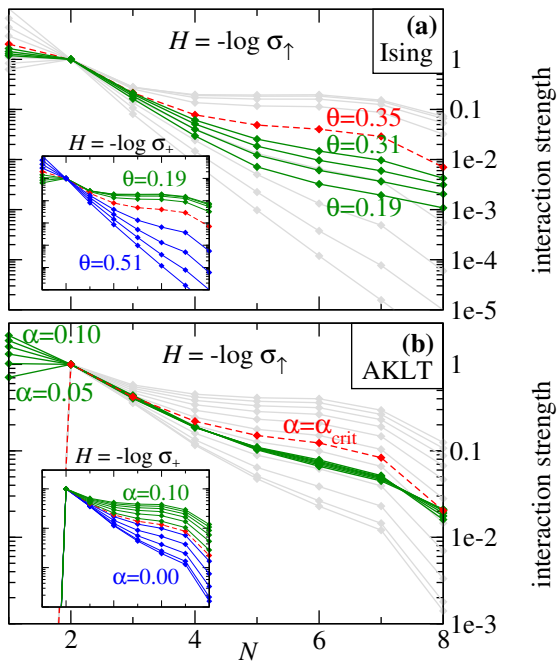}
\caption{
\label{fig:enthams}
Interaction strength vs.\ interaction range of 
$H_\mathrm{ent}$~\cite{cirac:peps-boundaries} for \textbf{(a)}
the Ising PEPS ($\theta=0.19,0.23,\dots,0.51$) and \textbf{(b)} the nematic
AKLT model ($\alpha=0,0.01,\dots,0.1$), where blue=trivial phase,
green=symmetry broken phase, and red=critical point. The insets show
$H_\mathrm{ent}$ computed from $\sigma_+$, where the interaction 
becomes long-ranged after the phase transition.  The main panels show
$H_\mathrm{ent}$ derived from the symmetry broken states (gray lines
indicate the data from $\sigma_+$), and we find that in both cases,
$H_\mathrm{ent}$ becomes more local again after the phase
transition.}
\end{figure}

Let us now turn towards entanglement spectra and the entanglement
Hamiltonian.  The entanglement spectrum of a state on e.g.\ an infinite
cylinder is given by the spectrum of the reduced density operator $\rho_L$
of the half-infinite cylinder, and its low-energy part can be interpreted
as the spectrum of a Gibbs state $e^{-H_\mathrm{ent}}$ of an ``entanglement
Hamiltonian'' $H_\mathrm{ent}$~\cite{li:es-qhe-sphere}.  In the context of
PEPS models, it has been shown~\cite{cirac:peps-boundaries} that the
spectrum of $\rho_L$ is equal to the spectrum of the symmetrized fixed
point of $\mathbb{T}$, 
\[
\sigma_\bullet= \sqrt{l_\bullet^T}r_\bullet\sqrt{l_\bullet^T}\ ,
\]
where $\bullet$ denotes a possible label of the fixed point, such as in the
case of multiple sectors.
In turn, $\sigma_\bullet$ allows to define the entanglement Hamiltonian 
through $H_\mathrm{ent}:=-\ln\sigma_\bullet$.  

In
Ref.~\cite{cirac:peps-boundaries}, $H_\mathrm{ent}$ has been studied 
for both the Ising PEPS and the nematic AKLT model based on the unique
fixed point of $\mathbb T$ for finite $N_v$, i.e., $\bullet=+$, and it has
been found that it is (quasi-)local (i.e., the interactions decay
exponentially with distance) in the trivial phase, the interaction length
diverges at the phase transition, and $H_\mathrm{ent}$ remains
long-ranged in the symmetry-broken phase; the corresponding results for
both models are shown  in the insets of Figs.~\ref{fig:enthams}a,b.

However, as we have argued, in the symmetry broken phase the
physically meaningful states, i.e., those which are stable under arbitrary
perturbations, are the symmetry broken ones, and we should therefore
rather use the symmetry broken fixed points $\bullet=\uparrow,\downarrow$
when determining $H_\mathrm{ent}$.  In Fig.~\ref{fig:enthams}, we compare
the interaction range of $H_\mathrm{ent}$ derived from 
$\bullet=\uparrow,\downarrow$~\footnote{%
Since for finite $N_v$, $r_++r_-$ is not positive, we have used
$\sigma_{\uparrow}\propto r_++(1+2\lambda)r_-$, with $\lambda =
1+\bra{\tau}r_-\ket{\tau}/\bra{\tau} r_+\ket{\tau}$.  Ideally, $\ket\tau$
should be choosen as the most negative eigenvector of $r_++r_-$, but we
have used $\ket\tau=\ket{1\dots1}$ which is a very good approximation to
it.
}
with the one obtained from $\bullet=+$ (insets) for the Ising PEPS
(Fig.~\ref{fig:enthams}a)~\footnote{%
Note that while the transfer operator of the Ising PEPS can be mapped to
free fermions and its fixed point is thus the ground state of a free
fermion Hamiltonian, it cannot be understood any more in terms of free
fermions after separating ket and bra index (as the corresponding map is
not Gaussian), and thus the entanglement
Hamiltonian of the Ising PEPS cannot be solved for analytically.}
and the nematic AKLT model (Fig.~\ref{fig:enthams}b):
We find that by considering the latter, the
locality of $H_\mathrm{ent}$ in the symmetry broken phase is restored,
and $H_\mathrm{ent}$ diverges only at the phase transition, in accordance
with the intuition that the interaction length of the entanglement
Hamiltonian should reflect the characteristic length scale of the system,
and thus should be finite away from the critical point.

Together with previous findings on how to recover locality of
$H_\mathrm{ent}$ in topological phases~\cite{schuch:topo-top}, this shows
that by considering the physically meaningful (this is, stable) fixed
points, a local entanglement Hamiltonian can be obtained for \emph{all}
gapped phases, as intuitively expected.

\section{Conclusions}

We have studied the occurence of long-range order and spontaneous symmetry
breaking in PEPS models.  We have shown that long-range order is closely
related to a degeneracy in the transfer operator, and have characterized
the symmetry broken (i.e., stable) states in terms of the fixed points
$r_\pm$ of the transfer operator, which we found to be the extremal
positive semi-definite states $r_{\uparrow\downarrow}=r_+\pm \lambda
r_-\ge0$ irrespective of the model and the perturbation.  These fixed
points do not only yield the physically relevant boundary conditions in
the symmetry broken phase, but each one by itself already carries the full
information about \emph{both} fixed points, as $r_\pm=r_\uparrow\pm
X^{\otimes N_v}r_\uparrow X^{\otimes N_v}$. Moreover, they give rise to
local entanglement Hamiltonians, thereby establishing locality of the
entanglement Hamiltonian for all gapped phases.

\begin{acknowledgments}
We acknowledge helpful conversations with 
Ignacio Cirac,
David Perez-Garcia, 
Volkher Scholz, 
and Frank Verstraete. 
This work has been supported by the Alexander von Humboldt foundation,
the ERC grant WASCOSYS, and by JARA-HPC through grants jara0092 and
jara0111.
\end{acknowledgments}

\vspace*{0.5cm}
\onecolumngrid
\hspace*{\fill}
\rule{12cm}{0.25mm}
\hspace*{\fill}

\appendix*

\section{Relation between long-range order and gap of the transfer
operator}

In this appendix, we give a detailed derivation of the equality of the
r.h.s.\ of Eqs.~(\ref{eq:lro-top}) and (\ref{eq:lro-top-fpt}).
We start by splitting
\[
\sum_{p=0}^{N_h-2}\frac{\tr[\mathbb{T}_{\hat Z}\mathbb{T}^p
		\mathbb T_{\hat Z} \mathbb T^{N_h-p-2}]}{
	\tr[\mathbb T^{N_h}]} = 
S(0,N_\mathrm{cut})
   + S(N_\mathrm{cut}+1,N_h-2)\ ,
\]
where $N_\mathrm{cut}=\lfloor \tfrac{N_h}{2}\rfloor-1$, and
\[
S(a,b):= \sum_{p=a}^{b}\frac{\tr[\mathbb{T}_{\hat Z}\mathbb{T}^p
		\mathbb T_{\hat Z} \mathbb T^{N_h-p-2}]}{
	\tr[\mathbb T^{N_h}]} \ .
\]
Due to cyclicity of the trace, 
$S(N_\mathrm{cut}+1,N_h-2)=
S(0,N_\mathrm{cut}+\kappa)$, where
$N_\mathrm{cut}+\kappa=N_h-2-(N_\mathrm{cut}+1)$ and thus $\kappa=-1,0$,
depending whether $N_h$ is even or odd.

We will now show that 
\[
\lim_{N_h\to\infty} S(0,N_\mathrm{cut}+\kappa) = 
\sum_{p=0}^\infty \tr\big[
	\mathbb T_{\hat Z} \mathbb T^p \mathbb T_{\hat Z}
	\rket{r_+}\rbra{l_+}\big]\ .
\]
To this end, we will make use of Eq.~(\ref{eq:top-exp-conv}),
\begin{equation}
\|\mathbb T^M-|r_+)(l_+|\,\|_{\mathrm{tr}}\le c \Gamma^{M}
    \tag{\ref{eq:top-exp-conv}}
\end{equation}
where $\Gamma<1$, as well as 
\begin{equation}
    \label{eq:app:TzTpTzX}
\big|\tr[\mathbb T_{\hat Z} \mathbb T^p \mathbb T_{\hat Z} X]\big|\le
\|\mathbb T_{\hat Z}\|_\mathrm{op}\,
\|\mathbb T^p\|_\mathrm{op}\,
\|\mathbb T_{\hat Z}\|_\mathrm{op}\,
\|X\|_\mathrm{tr} \le 
\zeta \|X\|_\mathrm{tr}
\end{equation}
with $\zeta := (c+1)\|\mathbb T_{\hat Z}\|^2_\mathrm{op}$, which can be shown using
H\"older's inequality, the submultiplicativity of the operator norm, and
\[
\|\mathbb T^p\|_\mathrm{op}\le
\big\|\mathbb T^p-|r_+)(l_+|\big\|_\mathrm{op}+
\big\||r_+)(l_+|\big\|_\mathrm{op}\le
\big\|\mathbb T^p-|r_+)(l_+|\big\|_\mathrm{tr}+1
\stackrel{(\ref{eq:top-exp-conv})}{\le}
c\Gamma^p+1\le c+1\ ,
\]
 and finally
\begin{equation}
    \label{eq:app:TN-lowerbnd}
\big|\tr[\mathbb T^{N_h}]\big|\ge
\big|\tr[|r_+)(l_+|]\big|-
\big|\tr[\mathbb T^{N_h}-|r_+)(l_+|]\big|
\ge 1 - \big\|\mathbb T^{N_h}-|r_+)(l_+|\big\|_\mathrm{tr}
\ge 1 - c\Gamma^{N_h}\ .
\end{equation}
We now have 
\begin{align*}
\Delta_p&:=
\left\lvert
\frac{\tr[\mathbb{T}_{\hat Z}\mathbb{T}^p
		\mathbb T_{\hat Z} \mathbb T^{N_h-p-2}]}{
	\tr[\mathbb T^{N_h}]}
-
\tr[\mathbb{T}_{\hat Z}\mathbb{T}^p
		\mathbb T_{\hat Z} |r_+)(l_+|]
\right\rvert
\\[1ex]
&\le 
\left\lvert
\frac{\tr[\mathbb{T}_{\hat Z}\mathbb{T}^p
		\mathbb T_{\hat Z} \mathbb T^{N_h-p-2}]}{
	\tr[\mathbb T^{N_h}]}
-
\frac{\tr[\mathbb{T}_{\hat Z}\mathbb{T}^p
		\mathbb T_{\hat Z} |r_+)(l_+|]}{
	\tr[\mathbb T^{N_h}]}
\right\rvert + 
\left\lvert
\frac{\tr[\mathbb{T}_{\hat Z}\mathbb{T}^p
		\mathbb T_{\hat Z} |r_+)(l_+|]}{
	\tr[\mathbb T^{N_h}]}
-
\frac{\tr[\mathbb{T}_{\hat Z}\mathbb{T}^p
		\mathbb T_{\hat Z} |r_+)(l_+|]}{
	\tr[|r_+)(l_+|]}
\right\rvert
\\[1ex]
&\stackrel{\mathclap{(\ref{eq:app:TzTpTzX},%
    \ref{eq:app:TN-lowerbnd})}}{\le}
\hspace*{1.3em}
\frac{\zeta\, \big\|\mathbb T^{N_h-p-2}-|r_+)(l_+|\big\|_\mathrm{tr}}{1-c\Gamma^{N_h}}
+ \zeta \big\||r_+)(l_+|\big\|_\mathrm{tr}\,
\left\lvert
\frac{\tr[|r_+)(l_+|]-\tr[\mathbb T^{N_h}]}{\tr[\mathbb T^{N_h}]\tr[|r_+)(l_+|]}
\right\rvert
\\[1ex]
&\stackrel{\mathclap{(\ref{eq:top-exp-conv},\ref{eq:app:TN-lowerbnd})}}{\le}
\hspace*{1.3em}
\frac{\zeta\, c\Gamma^{N_h-p-2}}{1-c\Gamma^{N_h}}
+
\zeta \big\||r_+)(l_+|\big\|_\mathrm{tr}\,
\frac{c\Gamma^{N_h}}{1-c\Gamma^{N_h}}
\\[1ex]
&\le 2 \zeta c\nu\Gamma^{N_h-p-2}\ ,
\end{align*}
where in the last step we have assumed that $N_h$ is sufficiently large such
that $1-c\Gamma^{N_h}\ge\tfrac12$, and have introduced
$\nu:=1+\big\||r_+)(l_+|\big\|_\mathrm{tr}$.
It follows that
\[
\left\lvert 
    S(0,N_\mathrm{cut}+\kappa)-
    \sum_{p=0}^{N_{\mathrm{cut}}+\kappa} \tr\big[
	\mathbb T_{\hat Z} \mathbb T^p \mathbb T_{\hat Z}
	\rket{r_+}\rbra{l_+}\big]
\right\rvert 
\le
\sum_{p=0}^{N_\mathrm{cut}+\kappa} \Delta_p
\le \tfrac{N_h}{2}\times 2\zeta c\nu \Gamma^{N_h/2-1}
\]
where we have used $N_{\mathrm{cut}}+\kappa\le \tfrac{N_h}{2}-1$ and
$N_h-p-2\ge N_h/2-1$. Clearly, the r.h.s.\ goes to zero as
$N_h\rightarrow\infty$, and thus,
\[
\lim_{N_h\to\infty} S(0,N_\mathrm{cut}+\kappa) = 
\sum_{p=0}^\infty \tr\big[
	\mathbb T_{\hat Z} \mathbb T^p \mathbb T_{\hat Z}
	\rket{r_+}\rbra{l_+}\big]
\]
as claimed.

\end{document}